\documentclass[aps,prx,reprint,superscriptaddress,longbibliography]{revtex4-2}

\usepackage{amsmath,amsfonts,amssymb}
\usepackage{bm}
\usepackage{physics}
\usepackage{xcolor}
\usepackage{graphicx}

\begin{document}

\title{Emergent Self-Attention from Astrocyte-Gated Associative Memory Dynamics}

\author{Arnau Vivet}
\affiliation{Departament d’Enginyeria Informàtica i Matemàtiques, Universitat Rovira i Virgili, 43007 Tarragona, Spain}

\author{Alex Arenas}
\affiliation{Departament d’Enginyeria Informàtica i Matemàtiques, Universitat Rovira i Virgili, 43007 Tarragona, Spain}
\affiliation{Complexity Science Hub Vienna, Metternichgasse 8, 1030 Vienna, Austria}

\date{\today}

\begin{abstract}
We introduce a Hopfield-type associative memory in which effective connectivity is multiplicatively modulated by astrocytic gains evolving under an entropy-regularized replicator equation. The coupled neuron–astrocyte dynamics admit a Lyapunov function, ensuring global convergence. At fixed points, astrocytic gains implement a softmax-normalized allocation over pattern similarity scores, yielding a mechanistic realization of self-attention as emergent routing on the gain simplex. In regimes of high memory load and interference, the model significantly improves retrieval accuracy relative to classical Hopfield dynamics and recent neuron–astrocyte baselines. These results establish a dynamical systems framework linking glial modulation, competitive resource allocation, and attention-like computation.
\end{abstract}

\maketitle

\section{Introduction}

Hopfield networks revolutionized our understanding of how collective neuronal dynamics implement 
associative memory with biological plausibility~\cite{Hopfield1982}. In their original formulation, 
neurons evolve under Hebbian connectivity that stores a finite set of patterns as energy minima. 
Although this framework provided a foundational bridge between statistical physics and neural 
computation, its practical applicability was limited by low storage capacity and the proliferation 
of spurious attractors under high pattern load or correlation~\cite{AmitBook1989}.

Interest resurged with generalized energy-based formulations, dense associative memory and modern 
Hopfield networks, that replace the quadratic Hebbian energy with higher-order or exponential 
interaction terms, dramatically expanding capacity and retrieval stability~\cite{Krotov2016, 
KrotovHopfield2020}. A key insight from this line of work is that the fixed-point 
readout of modern Hopfield networks is mathematically equivalent to the scaled dot-product attention 
mechanism at the core of Transformer architectures~\cite{Vaswani2017,Ramsauer2021}. This 
correspondence has reframed associative memory dynamics as a candidate substrate for attention and 
contextual reasoning in both artificial and biological systems.

Neuroscience has concurrently revealed that the brain's computational resources extend well beyond 
neurons. Astrocytes, glial cells long considered purely supportive, actively regulate synaptic 
transmission, integrate signals across local microcircuits, and contribute to memory-related 
plasticity~\cite{Araque1999,Perea2009,Letellier2016,DePitta2016,Giaume2010,VerkhratskyNedergaard2018}. 
Through Ca$^{2+}$ signaling, neurotransmitter uptake, and connexin-mediated gap-junction coupling, 
astrocytes modulate neuronal excitability and impose structured heterogeneity in effective synaptic 
strengths~\cite{Letellier2016,Giaume2010,VerkhratskyNedergaard2018}. 
Astrocytes have also been implicated in controlling circuit-level dynamical regimes, including the modulation of gamma-band synchronization associated with memory processing~\cite{purushotham2023astrocytic,makovkin2022controlling,thompson2021gamma}.
Critically, astrocytes are now 
recognized as active participants in memory encoding and consolidation: recent experimental work 
demonstrates that astrocyte ensembles selectively encode and stabilize salient experiences, with 
individual astrocytes reactivated during memory retrieval undergoing noradrenaline-dependent molecular 
tagging that promotes consolidation of specific traces~\cite{Dewa2025Nature,Zbaranska2025CellRes}. 
These findings extend the classical engram framework~\cite{Josselyn2020Science} by assigning 
astrocytes a mechanistic role in the selection and stabilization of memory representations, motivating 
their inclusion in computational models of associative memory.

Building on this foundation, recent theoretical work has proposed that astrocytic modulation could 
implement attention-like computation by providing a flexible, content-dependent reweighting of 
neuronal ensembles during retrieval~\cite{Kozachkov2023_BuildingTransformers, 
Kozachkov2025_NeuronAstrocyteAM}. In this view, self-attention corresponds to a gain reallocation 
across stored patterns driven by their relevance to the current query state. However, a precise 
dynamical account of how softmax-normalized gains arise as an emergent, self-organized property of 
neuron--astrocyte interaction, rather than being imposed by architectural design, has remained 
absent.

Here we close this gap by introducing a Hopfield-type associative memory in which effective 
connectivity is multiplicatively modulated by astrocytic gains that evolve on the probability simplex 
under an entropy-regularized replicator equation. The coupled neuron--astrocyte system admits a 
Lyapunov function, ensuring convergence to stationary points (equilibria) at which gains implement a 
Gibbs--Boltzmann (softmax) allocation over pattern similarity scores. The stationary readout is thus 
a convex combination of stored patterns with self-attention weights---emerging purely from the 
competitive resource dynamics of glial modulation, without any explicit attention mechanism being 
prescribed. Under high memory load and pattern interference, this mechanism substantially improves 
retrieval accuracy relative to classical Hopfield dynamics and recent neuron--astrocyte baselines, 
establishing a dynamical systems framework that links glial gain control, competitive resource 
allocation, and the emergence of attention-like computation in neural circuits.

\section{Astrocyte-Gated Associative Memory}

To ground the proposed astrocyte-neuron model, we begin by recalling the continuous-time classical Hopfield network, the minimal framework for associative memory. It is a class of recurrent neural network in which we consider a population of rate-based neurons evolving under a fully connected synaptic matrix. One of the key properties of this framework is that it can be derived from a Lyapunov function that guarantees convergence, enabling a precise characterization of memory retrieval as well as interference and capacity limits \cite{Hopfield1982,AmitBook1989}. For these reasons, the Hopfield model provides a natural substrate on which biologically motivated extensions can be constructed. 
More precisely, we consider $N$ rate units $x(t)\in\mathbb{R}^N$ with element-wise activation function $\phi(x) =\tanh(\sigma x)\in(-1,1)^N$ and synaptic weights $W\in\mathbb{R}^{N\times N}$:
\begin{equation}
\tau_x \dot x^i = - x^i + \sum_jW^{ij}\,\phi(x^j).
\label{eq:hopfield_rate}
\end{equation}
where $\tau_x$ is the neuronal rate time constant. It sets the intrinsic timescale over which the firing-rate state $x(t)$ relaxes toward the drive $W\,\phi(x)$ against the leak term. The synaptic matrix $W$ is constructed using Hebbian storage of $K$ binary patterns $\xi_\mu \in \{-1,1\}^{N}$ located at the vertices of the hypercube
\begin{equation}
W_{H}=\frac{1}{N}\sum_{\mu=1}^{K}\xi_\mu^i\xi_\mu^j.
\label{eq:hebb}
\end{equation}
Alternatively in matrix form we have $W_H= \frac{1}{N}\Xi \Xi^\intercal  $, where $\Xi \equiv (\xi_1,...,\xi_K) \in \{-1,1\}^{N\times K}$ is a binary matrix where patterns are stored as columns. Throughout the paper, we use superscripts to label neurons and subscripts to label memory patterns.

\subsection{Astrocyte-gated retrieval as pattern-wise gain control}

We extend the classical Hopfield rate dynamics by introducing $K$ astrocytic gain variables $p=(p_1,\dots,p_K)$ that modulate the contribution of each stored pattern to the synaptic matrix. The gains are constrained to the probability simplex,
\begin{equation}
p_\mu \ge 0 \quad (\forall \mu), \qquad \sum_{\mu=1}^{K} p_\mu = 1,
\end{equation}
so modulation is non-negative and globally conserved. The resulting neuronal dynamics read
\begin{align}
\tau_x \dot x^i
&= -x^i + \sum_{j=1}^{N} W^{ij}(p)\,\phi(x^j),
\label{eq:dx}
\end{align}
with effective synaptic matrix
\begin{equation}
W(p) \equiv \frac{K}{N}\sum_{\mu=1}^{K} p_\mu\, \xi_\mu \xi_\mu^\intercal
= \frac{K}{N}\,\Xi\,\mathrm{diag}(p)\,\Xi^\intercal .
\label{eq:astro_hebb}
\end{equation}
Thus, $p_\mu$ acts as a multiplicative gain on the $\mu$-th Hebbian outer product. The classical Hopfield coupling is recovered for uniform gains $p_\mu=1/K$, for which $W(p)=W_H$.

We model astrocytic modulation as an adaptive allocation process governed by a softmax-regularized replicator flow (hereafter ``s-replicator''):
\begin{align}
\tau_p \dot p_\mu
&= p_\mu\Big(F_\mu - \sum_{\nu=1}^{K} p_\nu F_\nu\Big),
\label{eq:dp}
\end{align}
which preserves the simplex constraints. The fitness is defined as
\begin{equation}
F_\mu \equiv f_\mu - T\log p_\mu, \qquad T>0,
\end{equation}
where
\begin{equation}
f_\mu \equiv \frac{1}{2N}\Big(\sum_{j=1}^{N} \xi_\mu^j \phi(x^j)\Big)^2
\end{equation}
is the squared overlap with pattern $\mu$, and the logarithmic term acts as an entropic (temperature-like) regularizer that discourages collapse of the gains onto a single pattern. Throughout, we use the same neuronal nonlinearity as in the classical model, $\phi(x^j)=\tanh(\sigma x^j)$ with $\sigma>0$.

We study retrieval dynamics by initializing the neuronal state $x(0)$ as a corrupted version of one of the stored patterns. Unless stated otherwise, the astrocytic gains are initialized uniformly, $p_\mu(0)=1/K$ for all $\mu$. During the dynamics, Eq.~\eqref{eq:dp} reallocates gain toward patterns with larger instantaneous overlap $f_\mu(x)$, thereby reshaping the effective coupling $W(p)$ in Eq.~\eqref{eq:astro_hebb}. This positive feedback selectively amplifies the synaptic contribution of the most compatible patterns and suppresses competing ones, reducing interference and effectively increasing the basin of attraction of the target memory.

The biophysical motivation for Eqs.~(\ref{eq:dx},\ref{eq:dp}) is the following. We view each stored pattern as an engram-like neuronal ensemble, and interpret the gain $p_\mu$ as an effective proxy for astrocyte-mediated modulation of the synaptic pathways supporting that ensemble. Specifically, $p_\mu$ summarizes (at a coarse-grained level) how local astrocytic activity can bias presynaptic release probability and thus rescale the effective strength of the synapses recruited by pattern $\mu$. This abstraction is inspired by evidence that synaptic activity can evoke highly localized astrocytic Ca$^{2+}$ signals in perisynaptic microdomains adjacent to individual synapses \cite{DiCastro2011NatNeurosci}. In several experimental settings, astrocytic Ca$^{2+}$ elevations have been reported to modulate presynaptic release probability and thereby reshape effective synaptic transmission, consistent with an activity-dependent astrocytic gating of synaptic pathways \cite{PereaAraque2007Science}. At the same time, the extent to which Ca$^{2+}$-dependent astrocytic signaling modulates excitatory transmission and plasticity under baseline conditions remains debated and depends on preparation and stimulation regime \cite{AgulhonFiaccoMcCarthy2010Science}; accordingly, we treat $p_\mu$ as a phenomenological variable that aggregates multiple microscopic mechanisms rather than as a direct readout of any single molecular process.

Finally, we constrain the astrocytic gains to the simplex, $p\in\Delta^{K-1}$, i.e.,
$p_\mu\ge 0$ and $\sum_{\mu=1}^K p_\mu=1$. In the model, this implements a finite modulatory capacity: increasing gain for one pattern necessarily reduces the gain available to others, thereby enforcing competitive allocation across pathways. This is a deliberate modeling device that operationalizes resource limitation and homeostatic competition at the level of pattern gains, rather than a claim that the brain explicitly maintains a globally conserved scalar ``budget'' \cite{ShigetomiPatelKhakh2016TrendsCellBiol}.

The neuron--astrocyte coupling is driven by a pattern-match functional $f_\mu(x)$, which quantifies how strongly the current neuronal state expresses pattern $\mu$ (via its overlap with $\xi_\mu$). We map this match to a fitness $F_\mu$ that controls the gain dynamics: patterns with larger $F_\mu$ are preferentially upweighted by Eq.~\eqref{eq:dp}. The additional logarithmic term in $F_\mu$ acts as an entropic regularizer, penalizing highly concentrated allocations and preventing winner-take-all behavior; consequently, the stationary distribution of gains is softmax-like, with sharpness set by the parameter $T$.

\subsection{Analytic results}
\label{sec:analytic_results}

We now summarize three analytic results that organize the theoretical picture and will be used throughout the remainder of the paper. First, we show that the coupled neuron--astrocyte dynamics admit a global Lyapunov function, so trajectories are dissipative and converge to the set of stationary points rather than exhibiting sustained oscillations or chaos. This structural result justifies focusing on equilibria. Second, we characterize these equilibria explicitly via coupled fixed-point equations for $(x^*,p^*)$, which reveal how astrocytic modulation implements a temperature-controlled, softmax-like allocation over patterns at stationarity. Finally, we connect the framework back to the classical Hopfield model by identifying limiting regimes in which the gain distribution becomes uniform and the effective synaptic matrix reduces to $W_H$. Together, these three results provide (i) a convergence guarantee, (ii) an interpretable description of the attractors, and (iii) a consistency link to the standard associative-memory baseline.

\subsubsection{The system is a global gradient flow}

A key property of the classical Hopfield network is the existence of an energy (Lyapunov) function that decreases along the trajectories thus guaranteeing convergence of the dynamics. As shown in the Appendix \ref{app:gradient_flow}, our system can also be written as a gradient flow whose energy is given by
\begin{equation} 
\begin{aligned}
\mathcal{L}(x,p) = &KT \sum_\mu p_\mu \log \frac{p_\mu }{\pi_\mu}-KT\log Z  \\
&+\sum_i^Nx^i\phi(x^i)  - L(x^i),
\end{aligned}
\label{eq:energy_index}
\end{equation}
where $\pi_\mu= \frac{1}{Z}e^{f_\mu/T}$ with $Z = \sum_\mu e^{f_\mu/T}$ and $L(x^i)=\frac{1}{\sigma }\log \cosh (\sigma \, x^i)$. The gradient flow condition implies that this quantity decreases along the trajectories 
$$\frac{d}{dt} \mathcal{L}(x,p;t)=-\|\nabla  \mathcal{L}(x,p;t)\|^2\leq 0  $$
. The last two terms of Eq.\ref{eq:energy_index} are similar to the modern Hopfield networks \cite{krotov2020large, Ramsauer2021}, while the first term is the potential for the s-replicator equation also encoding for neuron-astrocyte interaction. 
Since $\mathcal{L}$ is non-increasing and constant only at stationary points, the long-time behavior is fully determined by the coupled equilibria $(x^*,p^*)$, which we analyze next.

\subsubsection{Fixed point analysis}

The equilibrium $(x^*,p^*)$, satisfies the coupled fixed point condition
\begin{align}
\label{eq:fixed_point_eqs}
     &x^* = W(p^*) \,\phi(x^*) \\
   &p^*_\mu=\frac{\exp(f_\mu/T)}{\sum_\nu \exp(f_\nu/T)} = \operatorname{softmax}_\mu\!\left(\frac{f_\mu(x^*)}{T}\right).
\end{align}
This can be interpreted as a mixture of (rank 1) experts \cite{jacobs1991adaptive} commonly expressed as $x^*= \sum_\mu g_\mu (x^*) E_\mu (\phi(x^*))$, where the self-attention like routing $g_\mu(x) \equiv p_\mu $ emerges naturally from the s-replicator, and the experts are rank-1 matrices $E_\mu (\phi(x))\equiv \xi_\mu \xi_\mu^\intercal  \phi(x)$. In this interpretation, astrocytic modulation provides the routing mechanism that selects the relevant memory patterns in a context-dependent and dynamical manner. These equilibrium relations make clear that the model reduces to the classical Hopfield network whenever the gain distribution is forced to remain (or becomes) uniform, which we discuss next through two limiting regimes.

\subsubsection{Recovering the classical Hopfield model}

Our model recovers the classical associative memory in two dynamical regimes. For the first case we take the gain timescale to be infinite $\tau_p \to \infty$ such that the astrocytic influence remains constant for all time $\dot p_\mu=0$ (see Appendix \ref{app:fixed_points}). Since the modulation is initialized uniformly $p_\mu = 1/K$, we recover the classical Hopfield mechanism 
\begin{equation}
\tau_x \dot x = - x +W(1/K)\phi(x).
\label{eq:Hopfield_regime}
\end{equation}
This should be interpreted as the astrocytic processes being frozen, in which case there may not be a modulatory influence on the neuron dynamics. The second limit in which we recover the classical model is when we take $T\to \infty$, in which case the entropic term dominates and the only stable gain distribution is again uniform $p_\mu = 1/K$ (Appendix \ref{app:fixed_points}).

\subsection{Simulation results}

Our focus is on how two control knobs shape retrieval: (i) the relative timescales of neuronal relaxation and astrocytic routing, quantified by $\tau_x$ and $\tau_p$, and (ii) the selectivity parameter $T$, which sets the strength of entropic regularization in the gain dynamics. Because our dynamics are Lyapunov (Sec.~\ref{sec:analytic_results}), convergence is guaranteed in principle; numerically, however, we observe a competition between intrinsic convergence time and the finite simulation horizon $t_f$. We therefore report both end-point observables and empirical convergence times to disentangle genuine failure from slow convergence.
We emphasize that these simulations are intended as proof-of-principle demonstrations of dynamical regimes controlled by $(\tau_x,\tau_p,T)$.

\subsubsection{Dynamical analysis}

In this first analysis, the goal is to understand the long term behavior of the system. To that end, we define two observables, one representative of the neuron component and the other for the astrocytic gains. Our focus is on how two control knobs shape retrieval: (i) the relative timescales of neuronal relaxation and astrocytic routing, quantified by $\tau_x$ and $\tau_p$, and (ii) the selectivity parameter $T$, which sets the strength of entropic regularization in the gain dynamics. Because our dynamics are Lyapunov (Sec.~\ref{sec:analytic_results}), convergence is guaranteed in principle; numerically, however, we observe a competition between intrinsic convergence time and the finite simulation horizon $t_f$. We therefore report both end-point observables and empirical convergence times to disentangle genuine failure from slow convergence.
\\
We simulate a network of $N=30$ neurons storing $K=100$ binary (overlapping) patterns $\{\xi_\mu\}$ via Eq.~\ref{eq:astro_hebb}. To probe retrieval, we corrupt a reference memory $\xi_0 \to \xi_0^\eta $ by flipping $n$ randomly chosen entries, defining the noise level $\eta := n/N=0.2$ (six flipped bits). We initialize the neuronal state with the corrupted memory, $x(t_i)=\xi_0^\eta$ (see Appendix \ref{app:simulation_details}).

We quantify neuronal retrieval at $t_f$ using a \emph{soft} Hamming error,
\begin{equation}
\text{Error}=\epsilon(t_f) := \frac{1}{2}\sum_{i=1}^{N}\left|\xi_0^i-\phi(x^i(t_f))\right|,
\end{equation}
which reduces to the standard Hamming distance when $\phi(x^i)\in\{-1,1\}$ and satisfies $\epsilon_{\mathrm{soft}}\in[0,N]$.

To quantify how concentrated the gain vector $p(t_f)$ is over memories, we compute its Shannon entropy $H[p]$ and the associated perplexity 
\begin{equation}
P:=\exp(H[p]) = \exp(-\sum_{\mu=1}^{K} p_\mu \log p_\mu).
\end{equation}
Perplexity satisfies $P\in[1,K]$, with $P=1$ indicating near winner-take-all routing and $P\approx K$ indicating near-uniform allocation.
Unless stated otherwise, we set $\tau_x=1$, $\tau_p=1$, and $T=0.01$, and vary one parameter at a time.
\vspace{0.5\baselineskip}

\textit{Varying astrocyte timescale $\tau_p$ (keeping $\tau_x = 1$)} 
\vspace{0.5\baselineskip}

\begin{figure}[h]
    \centering
    \includegraphics[width=1\linewidth]{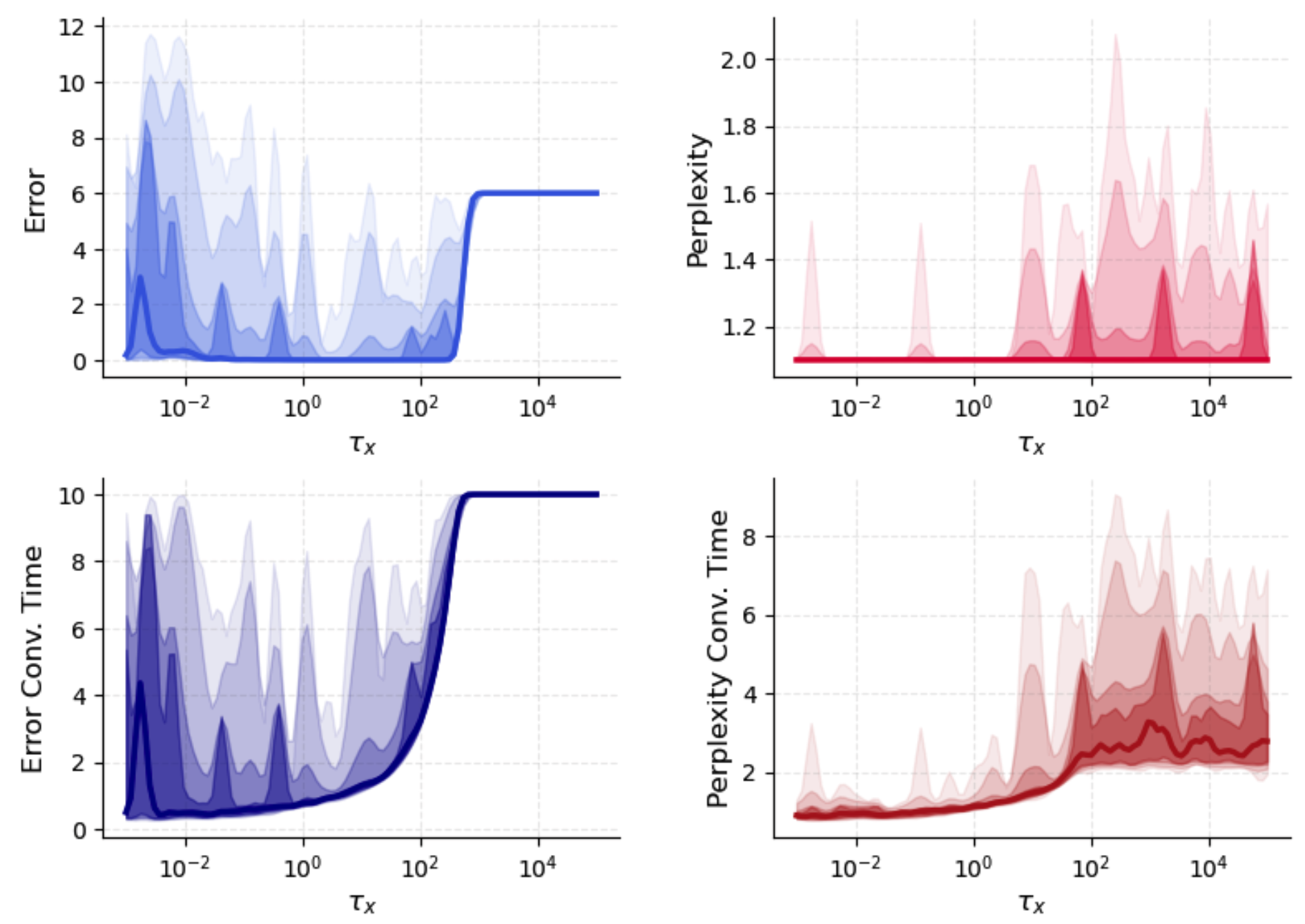}
    \caption{Final-time retrieval error $\epsilon(t_f)$ and gain perplexity $P(t_f)$ as functions of the neuronal timescale $\tau_x$ (with $\tau_p=1$), together with the corresponding convergence times. Solid lines show medians across trials; shaded bands denote percentile ranges $(5,95)$, $(10,90)$, $(20,80)$, and $(25,75)$.}
    \label{fig:x_timescale}
\end{figure}

Mechanistically, $\tau_p$ controls how rapidly the routing weights $p_\mu(t)$ track the instantaneous overlaps $f_\mu(x(t))$; small $\tau_p$ yields fast reweighting of $W(p)$ toward the correct pattern, whereas large $\tau_p$ delays this reshaping, i.e. $\tau_p$ parameter controls the gain $p(t)$ response time Fig.\ref{fig:x_timescale}. If we let astrocytic modulation be very fast $\tau_p \to 0$, the gains quickly concentrate on the most compatible memories, so the retrieval is fast and we get both low error and low perplexity. As $\tau_p $ increases, we see the gain adaptation become slower and the system needs more time to route the right memory. With a fixed simulation time, it looks like performance gets worse beyond $\tau_p \sim 10$, but this behavior is explained simply because the simulation is stopped before convergence. In the extreme $\tau_p\to \infty$ case, since the modulation essentially becomes frozen, the classical Hopfield regime is recovered, which performs poorly in this high memory storage regime (see also Appendix \ref{app:fixed_points}). 
\begin{figure}
    \centering
    \includegraphics[width=1\linewidth]{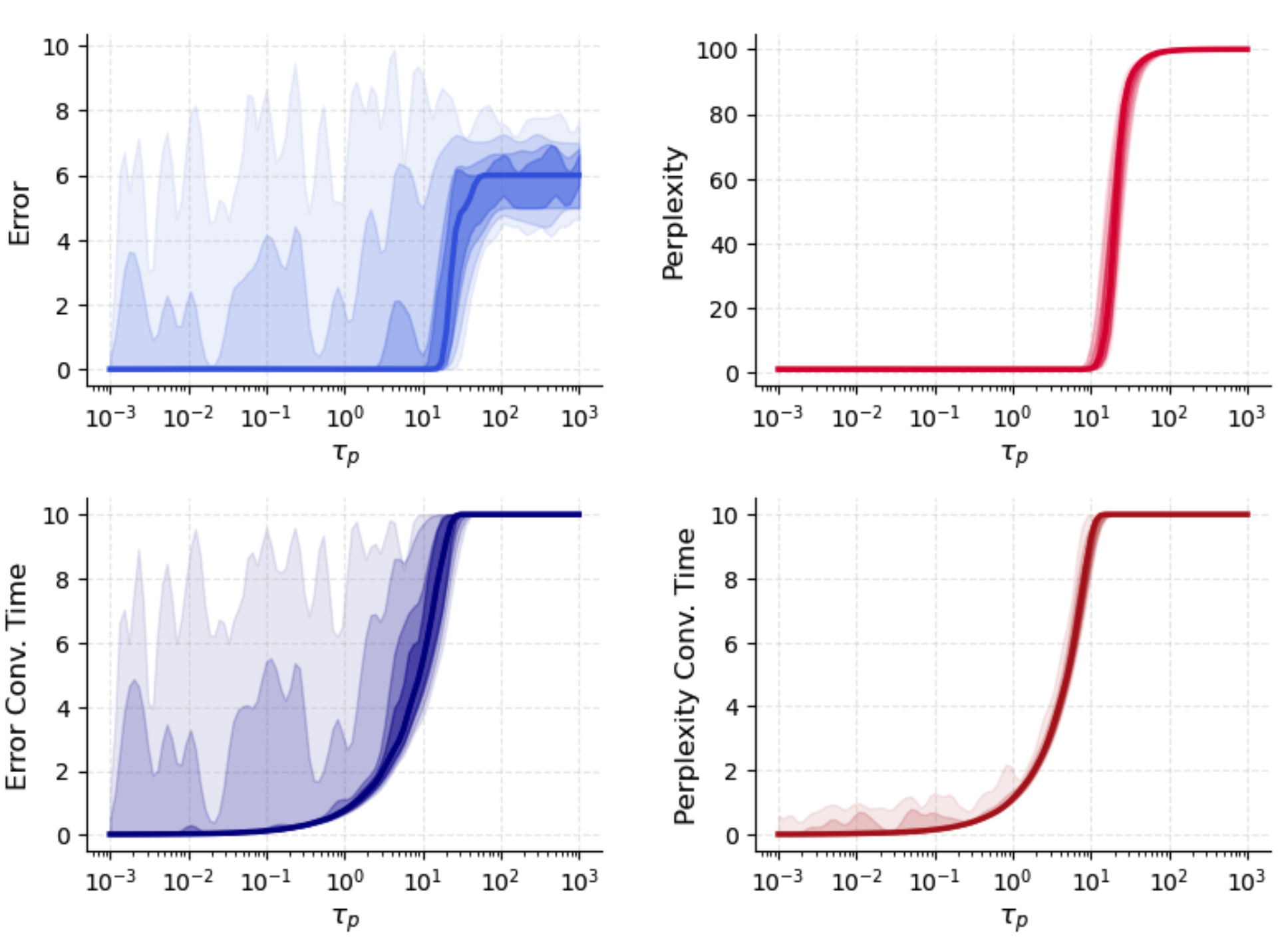}
    \caption{
    End-of-run retrieval error $\epsilon(t_f)$ and gain perplexity $P(t_f)$ as functions of the astrocytic timescale $\tau_p$ (with $\tau_x=1$), together with the corresponding convergence times. Solid lines show medians across trials; shaded bands denote percentile ranges $(5,95)$, $(10,90)$, $(20,80)$, and $(25,75)$.
    }
    \label{fig:p_timescale}
\end{figure}

\vspace{0.5\baselineskip}

\textit{Varying astrocyte timescale $\tau_x$ (keeping $\tau_p = 1$)} 

\vspace{0.5\baselineskip}

$\tau_x$ controls how fast neurons $x(t)$ relax Fig.\ref{fig:p_timescale}. Setting $\tau_x \to 0$ the neuron relaxation is essentially instantaneous, this makes the neuron configuration "commit" to an attractor too early, before the modulatory signal has time to reshape the landscape. Then the gain modifies the effective connectivity around the wrong attractor leading to high confidence (low perplexity) on the wrong memory. On the other hand, as we increase $\tau_x$, we see a decrease in the error as well as an increase in perplexity. High perplexity here means multiple patterns share similar overlap, so routing stays diffuse. This seemingly paradoxical region happens because as the neuron dynamics become slower, they do not have enough time to break the symmetry between patterns with similar fitness (see Appendix \ref{app:symmetry}). In fact, if we increase the time limit, symmetry breaks and we get high pattern selectivity again. If we keep increasing to $\tau_x \to \infty$, the convergence time increases super-linearly and we see the error stay exactly at $\epsilon (t_f)=6$ (in our simulations this happens at $\tau_x\sim 10^2$, simply because the simulation time is bounded) (see Appendix \ref{app:fixed_points}).

\vspace{0.5\baselineskip}

\begin{figure}
%    \centering
 \includegraphics[width=1\linewidth]{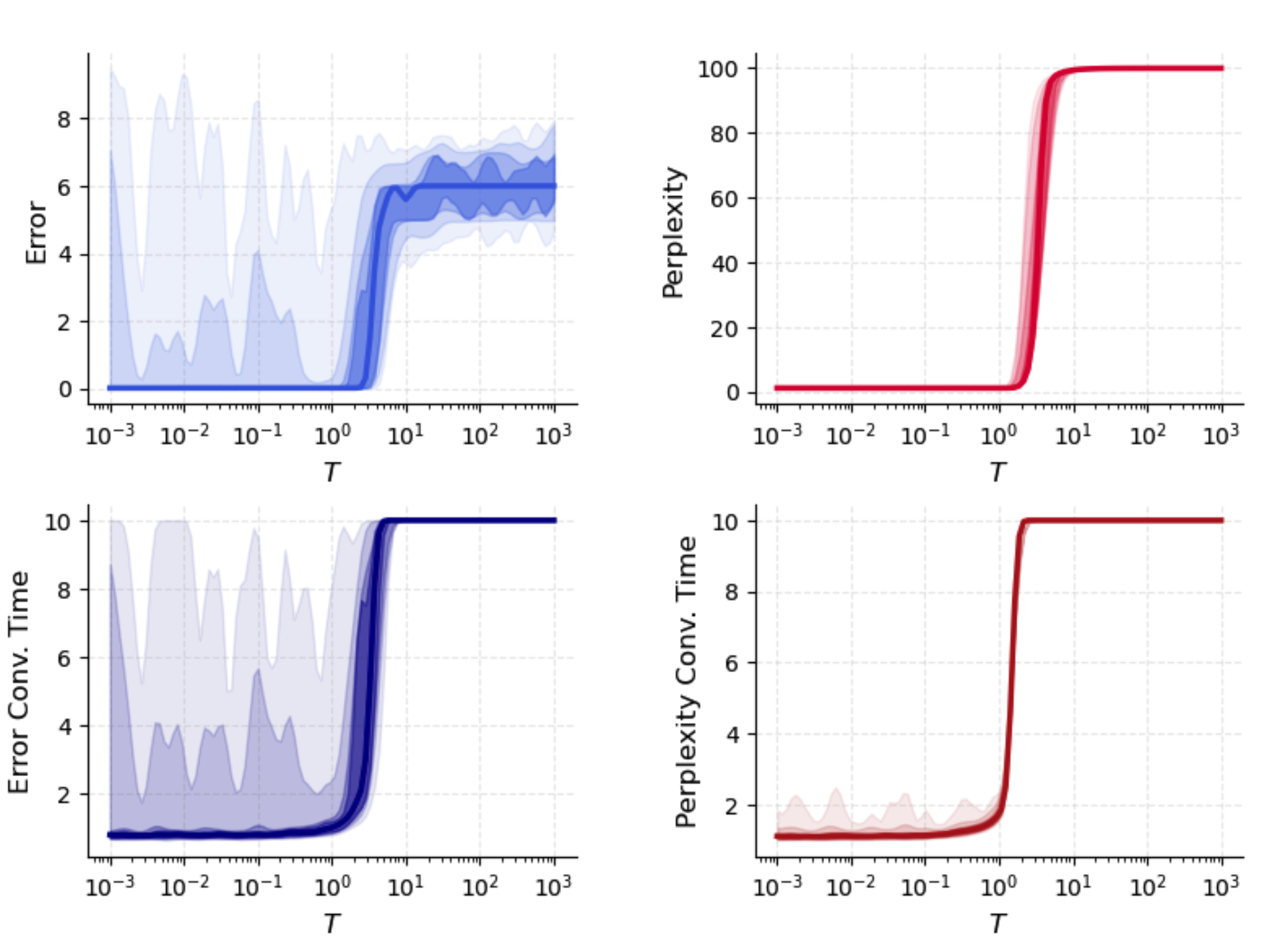}
    \caption{Final-time retrieval error $\epsilon(t_f)$ and gain perplexity $P(t_f)$ as functions of temperature $T$ (with $\tau_x=\tau_p=1$), together with the corresponding convergence times. Solid lines show medians across trials; shaded bands denote percentile ranges $(5,95)$, $(10,90)$, $(20,80)$, and $(25,75)$.}
    \label{fig:temperature_plot}
\end{figure}

\textit{Varying temperature $T$ (keeping $\tau_p =\tau_x = 1$):} 
\vspace{0.5\baselineskip}
Thus $T$ tunes a selectivity--robustness tradeoff: low $T$ yields sharp routing (small $P$) and effective interference suppression, whereas high $T$ enforces near-uniform gains and recovers Hopfield-like behavior, i.e. temperature controls astrocyte selectivity Fig.\ref{fig:temperature_plot}. When $T\to 0$, both error and perplexity are low since the regularization effect is weak and becomes highly selective. When $T\to \infty$ the error increases because the entropy dominates, meaning that $p$ stays close to uniform and once again we recover the Hopfield regime (see Appendix \ref{app:fixed_points}).

\subsubsection{Comparative retrieval performance}

\begin{figure*}[t]
    \centering
    \includegraphics[width=1\linewidth]{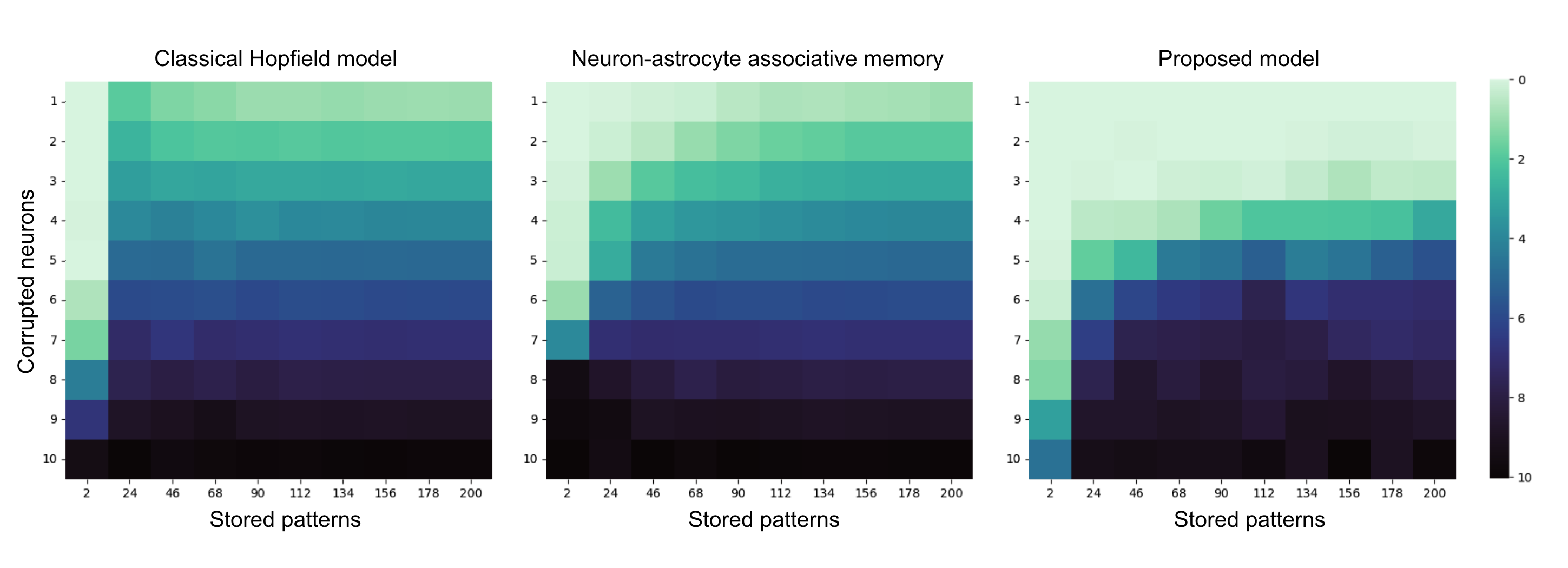}
    \caption{Retrieval benchmark across models. Each heat map reports the mean Hamming retrieval error $\epsilon(t_f)$ (lighter indicates lower error) as a function of memory load $K$ (x-axis) and corruption level $n$ (y-axis; number of flipped bits in the query pattern). Each entry is averaged over 50 independent random pattern realizations.}
    \label{fig:performance_comp}
\end{figure*}

We benchmark retrieval performance against (i) the classical Hopfield network and (ii) the neuron--astrocyte associative-memory model of Kozachkov \textit{et al.}~\cite{Kozachkov2025_NeuronAstrocyteAM}; see Fig.~\ref{fig:performance_comp}. To enable a dense scan over memory load and corruption, we set $N=20$ and vary the number of stored random binary patterns from $K=2$ to $K=200$. For each $(K,n)$ condition, we generate an independent pattern matrix $\Xi$, corrupt the target pattern $\xi_0$ by flipping $n$ randomly chosen entries (equivalently, corruption fraction $\eta=n/N$), and initialize the network as $x(t_i)=\xi_0^{(n)}$ (simulation protocol and model-specific parameters are reported in Appendix~\ref{app:simulation_details}). All models are evaluated under the same initialization and the same finite simulation horizon $t_f$.

We quantify retrieval at time $t_f$ by binarizing the final state and computing the Hamming error with respect to the target:
\begin{equation}
\epsilon(t_f) = \frac{1}{2}\sum_{i=1}^{N}\left|\xi_0^i-\mathrm{sign}\!\big(x^i(t_f)\big)\right|,
\end{equation}
so that $\epsilon\in[0,N]$ and $\epsilon=0$ indicates perfect retrieval. For statistical stability, we repeat each condition 50 times with independently sampled $\Xi$ and report the mean error.

Across the tested $(K,n)$ grid, our model yields lower mean retrieval error than both baselines, with the largest gains appearing at high memory loads where interference is strongest (Fig.~\ref{fig:performance_comp}).

\section{Discussion}
We introduced an astrocyte-gated extension of a Hopfield-type associative memory in which astrocytic processes allocate a finite modulatory capacity across stored patterns. By dynamically reweighting pattern-specific contributions to the effective connectivity, this additional degree of freedom reduces interference during retrieval and enlarges the regime in which accurate recall is achieved at high memory load.

We interpret each stored pattern as a memory-specific (engram-like) neuronal ensemble and the gain variable $p_\mu$ as a coarse-grained proxy for the effective strength of astrocyte-mediated modulation of the synaptic pathway supporting pattern $\mu$. This abstraction is motivated by the tripartite-synapse framework, in which astrocytes integrate local activity and can modulate synaptic efficacy through Ca$^{2+}$-dependent signaling and related pathways \cite{Araque1999,Perea2009,Letellier2016,DePitta2016,Giaume2010,VerkhratskyNedergaard2018}. Importantly, we do not interpret $p_\mu$ as a direct readout of any single molecular mechanism; rather, it summarizes pathway-level modulation at the scale relevant for associative retrieval.

A central modeling assumption is the simplex constraint $\sum_{\mu} p_\mu=1$, which enforces competitive allocation: increasing gain for one pathway necessarily reduces gain available to others. This operationalizes finite modulatory capacity (e.g., limited signaling/metabolic resources distributed across microdomains) and homeostatic competition at the level of pattern gains, without implying that the brain literally implements a globally conserved scalar ``budget.''

Mechanistically, the gain dynamics implement state-dependent competition. When the current neuronal state is more compatible with pattern $\mu$, the corresponding match $f_\mu(x)$ increases the fitness $F_\mu$ and thereby upweights $p_\mu$, which amplifies the $\mu$-th contribution to $W(p)$ while suppressing competitors. The logarithmic term in $F_\mu$ acts as an entropic regularizer: it penalizes highly concentrated allocations and discourages winner-take-all routing. In this view, the temperature parameter $T$ controls selectivity: low $T$ yields sharp, concentrated routing, whereas high $T$ enforces broader, near-uniform modulation and recovers Hopfield-like behavior. The joint neuron--astrocyte system remains Lyapunov-consistent (it admits a global Lyapunov function), ensuring convergence of the coupled dynamics.

Modern Hopfield networks and related dense-memory models achieve softmax-like retrieval through neuronal energy descent shaped by higher-order or exponential storage functions \cite{Krotov2016,KrotovHopfield2020,Ramsauer2021,Kozachkov2023_BuildingTransformers,Kozachkov2025_NeuronAstrocyteAM}. Our model produces a similar normalized reweighting, but via a different mechanism: normalization arises from competitive dynamics on the gain simplex (replicator-type routing) rather than being hard-wired into the neuronal energy landscape. In regimes where gains adapt rapidly, the stationary allocation approaches
\[
p^*_\mu \propto \exp\!\left(\frac{f_\mu}{T}\right),
\]
so retrieval can be viewed as operating with a dynamically reweighted subset of memories. Compared to the classical Hopfield model, where high load is associated with numerous spurious mixture states, our routing variable tends to concentrate weight on a small set of candidates under ambiguity, consistent with the low-perplexity regimes observed in simulations.

In addition, the degree to which attention-like structure is explicit in our framework depends on the choice of pattern score. While the main model uses a squared-overlap score $f_\mu(x)\propto \langle \xi_\mu,\phi(x)\rangle^2$ (hence invariant under $\langle \xi_\mu,\phi(x)\rangle\mapsto -\langle \xi_\mu,\phi(x)\rangle$), one may alternatively define a linear-overlap score $f_\mu(x)=\langle \xi_\mu,\phi(x)\rangle$. In this variant, the neuronal drive becomes a linear combination of stored patterns weighted by the gain vector, i.e. the update takes the form $\dot x = -x + \Xi p$ (up to timescale factors), and in the fast-gain limit the stationary allocation becomes $p^*\propto \exp(\Xi^\intercal \phi(x)/T)$, yielding the standard attention readout $x \leftarrow \Xi\,\mathrm{softmax}(\Xi^\intercal\phi(x)/T)$ \cite{Vaswani2017,Ramsauer2021}. We do not analyze this variant further here; we include it to emphasize that competitive routing on the gain simplex can mechanistically reproduce attention-like computation under closely related definitions of pattern compatibility.

Numerically, this modulatory degree of freedom improves retrieval accuracy across increasing memory load $K$ and increasing corruption level $n$ (equivalently $\eta=n/N$), outperforming both the classical Hopfield model \cite{Hopfield1982} and the neuron--astrocyte associative-memory baseline \cite{Kozachkov2025_NeuronAstrocyteAM} in the tested regimes (Fig.~\ref{fig:performance_comp}). Intuitively, adaptive gain allocation reshapes the effective connectivity (and hence the attractor structure) during recall by amplifying task-relevant patterns and suppressing competitors, in line with recent evidence that time-dependent modulation can improve robustness of Hopfield-type retrieval \cite{Betteti2025InputDriven}.

The model also suggests qualitative dependencies that could guide future theoretical and computational work. In particular, changes in astrocytic kinetics or background modulatory tone would be expected to alter the sharpness of pattern selection (captured here by $T$), while retrieval speed and stability should depend on the ratio $\tau_p/\tau_x$ setting the competition between routing and neuronal relaxation. More generally, the framework highlights a route by which transient changes in astrocyte-mediated signaling could bias recall without invoking synaptic plasticity. A limitation of the present formulation is that $p_\mu$ aggregates multiple biophysical processes and timescales into a single effective variable; identifying which cellular mechanisms and dynamical regimes can realize competitive allocation of this type remains an important direction for future work, and connects naturally to multi-timescale theories of memory stabilization \cite{BennaFusi2016}.

An important direction is to convert the proposed competitive routing mechanism into a scalable ML architecture, where $p$ acts as a parameterized, context-dependent gating distribution over memory components and $T$ controls gating entropy. This suggests lightweight mixture-of-experts memories with explicit competition and separable routing and state-update timescales (set by $\tau_p/\tau_x$). Developing stable, differentiable implementations and benchmarking them on noisy retrieval and continual-learning tasks are natural next steps.

\begin{acknowledgments}
We thank Prof. Luiz Pessoa and Prof. Sergio G{\'o}mez for discussions on astrocyte kinetics and associative memory. This work has been supported by spanish Ministerio de Ministerio de Ciencia, Innovación y Universidades PID2024-158120NB-C21. AA also acknowledges the ICREA Academia program of Generalitat de Catalunya.    
\end{acknowledgments}

\appendix
\section*{Appendix}

For the reader’s convenience, we collect here the few definitions that are repeatedly used in the derivations below, so the proofs can be followed without having to refer back to the main text. In particular, the effective synaptic matrix is $W(p)=\frac{K}{N}\,\Xi\,\mathrm{diag}(p)\,\Xi^\intercal
=\frac{K}{N}\sum_{\mu=1}^K p_\mu\,\xi_\mu\xi_\mu^\intercal$.
For the gain dynamics, $p\in\Delta^{K-1}$ implies $\mathbf{1}^\top \dot p = 0$ (tangent-space constraint), and the projector acts on $T_p\Delta^{K-1}$. In the gradient-flow formulation we use the Shahshahani (Fisher information) metric on $\Delta^{K-1}$, $G(p)=\mathrm{diag}(1/p)$, and the diagonal metric on neuronal pre-activations, $H_x=\mathrm{diag}(\phi'(x))$ with $\phi'(x)=\sigma\,\mathrm{sech}^2(\sigma x)$ for $\phi(x)=\tanh(\sigma x)$.

\subsection{Gradient-flow structure and Lyapunov function}
\label{app:gradient_flow}

Here we show that the coupled dynamics in Eqs.~(\ref{eq:dx},\ref{eq:dp}) admit a Lyapunov function and can be written as a (projected) gradient flow. We divide the discussion into the astrocyte domain and the neuron domain, where we derive their respective potentials.
Note that the coupled dynamics occur in $x,p \in \mathbb{R}^N\times\Delta^{K-1}$ and the metric (needed for the gradient flow result) is given by a block-diagonal (direct-sum) metric
\[
G_{x,p} = H_x\oplus G_p.
\]
This metric is composed of the astrocyte and neuron domain metrics. The former is given by the Fisher information metric (equivalently the Shahshahani metric), whose components are
\[
 g_{\mu\mu'}(p)=\frac{\delta_{\mu\mu'}}{p_\mu},
\]
so in matrix form $G(p)=\mathrm{diag}(1/p)$. The latter is given by 
\[
H_x = \mathrm{diag}(\phi'(x)),
\]
where in our case $\phi'(x) = \sigma sech^2(\sigma x)$.

\subsubsection{Astrocyte domain}
\label{app:astrocyte_domain}

For the astrocyte domain, we first show how the simplex constraints are imposed via the orthogonal projection and then we define the s-replicator potential. With this, we obtain the ODE for the $p$ variable and we discuss its fixed points. 

Following \cite{mertikopoulos2018riemanniangamedynamics}, given that we have a differential 0-form (or scalar potential) $\Phi:\mathbb{R}^K\to \mathbb{R}$ and a metric $G(p)$, we will define an ODE by computing the gradient of the scalar potential, and then projecting it onto the tangent space of the simplex $\dot p \in T_p\Delta^{K-1} $ (i.e. $\mathbf{1}^\intercal \dot p = \sum_\mu^K \dot p_\mu =0$). \\
In more detail, to compute the gradient, we get the 1-form which lives in the cotangent space $d\Phi \in T_p^* \mathbb{R}^K$, since the gradient lives in the tangent space, using the sharp map musical isomorphism $\#_G:T^*_p \mathbb{R}^K\to T_p \mathbb{R}^K$, the gradient becomes $z = G^{-1}(p)d\Phi (p)$ where $z \in T_p\mathbb{R} ^K$. \\
To compute the projection onto the tangent space of the simplex $T_p\Delta^{K-1}$ (since we want the evolution law of the ODE to remain on the simplex), we use the orthogonal projector map $\Pi_G: T_p\mathbb{R}^K \to T_p \Delta^{K-1}$, which takes $z$ to the closest vector in $T_p \Delta ^{K-1}$. That is, given a vector $z$, we solve for $\Pi_G(z) = {\arg \min}_{\dot p \in T_p\Delta^{K-1}} ||\dot p - z||_G^2$, which returns the vector in $T_p\Delta^{K-1}$ satisfying the minimum distance condition. Written in variational form
\[ \mathcal{L}(\dot p,\lambda) =  \frac{1}{2} ||\dot p-z||^2_G + \lambda \mathbf{1}^\intercal \dot p,\]
its minimum corresponds to the projected vector. Thus, solving for the minimum condition $\nabla_{\dot p} \mathcal{L} =0= G(\dot p - z) +\lambda \mathbf{1}  =0$, we get that:
\begin{equation} \dot p = z+\lambda G^{-1}\mathbf{1} .
\label{eq:minimum_proj}
\end{equation}
The condition of a tangent vector to the simplex is $\mathbf{1}^\intercal \dot p = 0$, we obtain \begin{equation}
\lambda=\frac{\mathbf{1}^\intercal z}{\mathbf{1}^\intercal G^{-1}\mathbf{1}}.
\end{equation}
where $\mathbf{1}^\intercal G^{-1}\mathbf{1} = \sum_i p_i=1$. Plugging $\lambda$ into Eq. (\ref{eq:minimum_proj}) and using $z = G^{-1}d\Phi$:
$$ \dot p  =[\mathbb{I}-G^{-1} \mathbf{1} \mathbf{1}^\intercal]G^{-1} d\Phi,$$
and developing the expression, the ODE becomes
\begin{equation} \label{eq:rep_gf}
\dot p = (\text{diag}(p)-pp^\intercal ) d\Phi .
\end{equation}

Now that we have our gradient flow ODE, we choose our $0$-form to be $\Phi (p)  = TD_{KL}(p||\pi)$ with $ \pi = \frac{1}{Z} e^{\frac{1}{T}f }$, such that  
\[\label{eq:free_energy}\Phi (p) =-\langle  p,F\rangle +T\log Z,\]
where $F = f-T\log p$ and the differential is given by $d\Phi = T\log p+T\mathbf{1}-f$ (where neither $\log Z$ nor $f$ depend on $p$). Substituting into Eq.\ref{eq:rep_gf}, we can write the gradient flow as
\begin{equation} \label{eq:replicator}
\dot p =- (\text{diag}(p)-pp^\intercal ) (T\log p-f), 
\end{equation}
where $T \mathbf{1}$ is proportional to $\mathbf{1}$ and $(\text{diag}(p)-pp^\intercal)\mathbf{1} =0$. We refer to this equation as the\textit{ s-replicator} (for "soft" replicator) as the entropic term $T\log p$ prevents winner-take-all dynamics for $T>0$. 

The fixed points of Eq.\ref{eq:replicator} satisfy
\[ 0 = (\text{diag}(p)-pp^\intercal ) F ,\]
Where $F=f-T\log p$. One (degenerate) way to satisfy this condition is to make $(\text{diag}(p)-pp^\intercal )=0$ where we get two conditions $p_\mu = p_\mu ^2$ and $p_\mu p_\nu = 0$. It can only be satisfied when $p$ is a one-hot vector, thus, the solution lies on the boundary (vertices) of the simplex. Solutions inside the simplex (where $p_\mu >0 \,\,\forall \mu$) satisfy

\[\text{diag}(p)F =pp^\intercal F.\]
Expressing it element-wise, since $p_\mu >0$, dividing by $p_\mu$
\[ 0=    F_\mu  -  \sum_\nu^K p_\nu F_\nu \quad \forall \mu ,\]
using that $F_\mu=f_\mu-T\log p_\mu$ and solving for $p_\mu$
\begin{equation} \label{eq:intermediate_p}
\log p_\mu = \frac{1}{T } \left[ f_\mu -  \sum_\nu^K p_\nu F_\nu \right]\Rightarrow  p_{\mu } = e^{ \frac{1}{T }  \left[ f_\mu -  \sum_\nu^K p_\nu F_\nu \right]}.
\end{equation}
Imposing $\sum_\mu ^Kp_\mu = 1$ gives
\[1= e^{- \frac{1}{T }  \sum_\nu^K p_\nu F_\nu }\sum_\mu^K e^{ \frac{1}{T }  f_\mu },\]
where taking logarithms, we can identify $ { \sum_\nu p_\nu F_\nu }=T\log \sum_\mu e^{ \frac{1}{T }  f_\mu } $ and we can recognize the partition function $Z =\sum_\mu e^{ \frac{1}{T }  f_\mu }$. Finally we can rewrite Eq.\ref{eq:intermediate_p} as
\begin{equation}
\label{eq:rep_fp}
    p_{\mu } = \frac{1}{Z}e^{ \frac{1}{T }  f_\mu },
\end{equation}
where in vector form $  p^* = \text{softmax}\left(\frac{1}{T}f^* \right) $. 

\subsubsection{Neuron domain}
\label{app:neuron_domain}

Here we provide a derivation of the classical Hopfield dynamics starting from the energy. We see that the metric $H_x = \text{diag}(\phi'(x))$ emerges as a natural choice. The Hopfield model energy \cite{Hopfield1982} is a scalar function $E: \mathbb{R}^N\to \mathbb{R}$ expressed as 
\begin{equation}\label{eq:hopfield_energy}
E(x) = -\frac{1}{2}\sum_{ij}^N\phi(x^i)W^{ij}\phi(x^j) + \sum_i^N [ x^i\phi(x^i)  - L(x^i)],
\end{equation}
where $L(x^i)=\int^{x^i} \phi(u)du $ and $\phi(x) = \tanh(k\, x)$, thus $L(x^i) = \frac{1}{k} \log \cosh (\sigma \, x^i)$. It can be shown that this potential induces a global gradient flow similarly to \cite{halder2019hopfieldneuralnetworkflow}. To show this, we compute the differential of the energy $dE = \sum_i \frac{\partial E}{\partial x^i}dx^i$ lying in $dE \in T_x^* \mathbb{R}^ N$ and convert it to a gradient flow using the sharp map just like before using the metric $H_x = \text{diag}(\phi')$. Differentiating the first term $E_1$, we get
\begin{equation}
\begin{aligned}
\partial_{x^k} E_1 &= -\sum_{ij} [\partial^{x^k} \phi(x^i) W^{ij} \phi(x^j)+ \phi(x^i) W^{ij}\partial_{x^k} \phi(x^j) ]\\
&=  -\sum_i \phi(x^i)W^{ki}\phi'(x^k),
\end{aligned}
\end{equation}
where we have assumed that $W=W^\intercal $. For the second term $E_2$, we get:
\begin{equation}\label{eq:neuron_term}
    \partial_{x^k } E_2 =x^k\phi'(x^k).
\end{equation}
Combining both terms, 
\[ \partial_{x^k}E=  \left[x^k-\sum_iW^{ik}\phi(x^i)\right]\, \phi'(x^k),\]
or in vector form 
\[dE = \text{diag}(\phi')[x-W\phi] .\]
Taking the sharp map $\#_H:T^*_x \mathbb{R}^N\to T_x \mathbb{R}^N$, we get $\dot x = -H^{-1}dE$, where the metric necessarily becomes
\[H_x = \text{diag}(\phi'(x)).\]
Note that $1/\phi'(x)>0$ since in our case $\phi' = sech^2$ . Because $H_x$ is diagonal, it also satisfies $h_x(u,u)=0 $ if $u=0$ and $h_x(u,v) = h_x(v,u)$ for all $x$. This is exactly the same as \cite{halder2019hopfieldneuralnetworkflow} up to the change of variable $ z = \phi(x)$, where $z$ is their dynamical variable, this way, $\dot { z} = \phi'(\phi^{-1}(z))[\phi^{-1}(z)-Wz]$ is exactly $\phi' \dot x = \phi' [x-W\phi]$.

\subsubsection{Joint potential}

Finally, we can combine the parts to define the joint potential function $\mathcal{L}:\mathbb{R}^N\times \Delta^K \rightarrow \mathbb{R}$ for the whole system from which we can derive the dynamics as a gradient flow. We define it as
\[ \mathcal{L}(x,p) = -KT\log Z+KTD_{KL}(p||\pi)  +[\langle x,\phi\rangle  - L(x)],\]
which combines the two potentials we have described so far. From Eq.\ref{eq:free_energy}, we know that $TD_{KL}(p||\pi) = -\langle p,F\rangle+T\log Z$ and we can express it like
\[ \mathcal{L}(x,p) =     -K\langle  p, f\rangle+KT\langle p,\log p\rangle +[\langle x,\phi\rangle  - L(x)],\]
where $ f_\mu = \frac{1}{2N} \left(\sum_i  \xi^i_\mu \phi^i\right)^2 $. In this form, the interaction term $\langle p,f\rangle = \frac{1}{2N} \sum_\mu p_\mu\langle \xi_\mu ,\phi\rangle^2 $ is in direct analogy to $\frac{1}{2}\phi W\phi = \frac{1}{2N}\sum_\mu \langle \xi_\mu,\phi\rangle ^2 $ of Eq.\ref{eq:hopfield_energy}, where now each term comes multiplied by the modulation $p_\mu $. The second term is the astrocytic regularization and the last one is the neuron activation term. 
From this potential we can derive the dynamics of our system by taking the differential living in $d\mathcal{L}(x,p)\in T^*_{x,p} (\mathbb{R}^N\times  \Delta^{K-1})$ such that: 
\begin{equation} \label{eq:differential_full}
d\mathcal{L}(x,p) = \sum_i^N \frac{\partial \mathcal{L}}{\partial x^i } dx^i  +\sum_\mu ^K \frac{\partial \mathcal{L}}{\partial p_\mu }dp_\mu. 
\end{equation}
For the $\partial_p \mathcal{L}(x,p)$ term, it is exactly the same as in Eq.\ref{eq:replicator}. For the first derivative, we have the interaction and activation terms. The activation term is exactly the same as in Eq.\ref{eq:neuron_term}, and for the interaction term we have
\[K\partial_{x^k} \langle p,f\rangle = \frac{K}{N} \sum_{\mu}^K\sum_i ^N   p_\mu \xi_\mu^i \xi_\mu^k \phi(x^i)  \phi'(x^k),\]
where in vector form $\frac{K}{N}\sum_\mu \xi_\mu^k p_\mu\langle \xi_\mu,\phi(x)\rangle \phi'(x^k) =\text{diag}(\phi')  W(p)  \phi(x) $, where we define the weight matrix as: 
\[W(p) = \frac{K}{N}\Xi^\intercal \text{diag}(p) \Xi\]
Putting everything together, the two terms of Eq.\ref{eq:differential_full} become:
\begin{align}\label{eq:dyn_equations}
    &\partial_x\mathcal{L}(x,p) =  \text{diag}(\phi')[x-W(p)\phi]  \\
    &\partial_p\mathcal{L}(x,p) =K(T\log p+T\mathbf{1}-f ).
\end{align}
Finally, to get the gradient flow, the metric of the whole space is the direct sum $M(x,p) = H(x)\oplus G(p)$ and the orthogonal projection acts only on the astrocyte subspace, thus $\Pi =\mathbb{I}\oplus  \Pi_G = \mathbb{I}\oplus (\mathbb{I}-G^{-1}\mathbf{1}^\intercal \mathbf{1}G^{-1})$. The natural gradient descent of $\mathcal{L}(x,p)$ can finally be expressed as
\[\begin{pmatrix}
   \tau_x \dot x\\
    \tau_p\dot p
\end{pmatrix} = -\begin{pmatrix}
\mathbb{I} &0\\
0   &  \,\Pi_G  
\end{pmatrix}
\begin{pmatrix}
H^{-1} &0\\
0   &  \,G^{-1}  
\end{pmatrix} \begin{pmatrix}
    \partial_x \mathcal{L}(x,z)\\
    \partial _p \mathcal{L}(x,p)
\end{pmatrix}. \]
Thus, the dynamics are as intended 
\begin{align} \label{eq:system_dynamics}
    &\tau_x \dot x= -x+W(p) \phi(x)\\
    &\tau_p \dot p =  (\text{diag} (p) - p p^\intercal) (f-T\log p), 
\end{align}
where we renormalize the time constant $\tau_p/K \rightarrow \tau_p $. This completes the proof that the system is a gradient flow by construction. 

\subsection{Dynamical analysis of the system:}

Now that we have defined the evolution of the system, we can explore both its symmetries as well as its long-term behavior in different parameter regimes to better understand its dynamics.

\subsection{Symmetry remarks}
\label{app:symmetry}

\subsubsection{$\mathbb{Z}_2$ invariance of the squared-overlap score}
In the main model the pattern score is a squared overlap,
$f_\mu(x)\propto \langle \xi_\mu,\phi(x)\rangle^2$.
Consequently, the gain dynamics are invariant under the sign flip
$\langle \xi_\mu,\phi(x)\rangle\mapsto -\langle \xi_\mu,\phi(x)\rangle$.
Equivalently, defining $m_\mu(x)\equiv \langle \xi_\mu,\phi(x)\rangle$, the gain update depends only on $m_\mu^2$ and cannot distinguish $m_\mu$ from $-m_\mu$. More generally, if two patterns satisfy $f_\mu(x)=f_\rho(x)$ at a given state, the gain update has no instantaneous preference between them.

\subsubsection{Symmetry breaking by neuronal dynamics}
Even with the squared-overlap score, degeneracies such as $f_\mu(x)=f_\rho(x)$ are typically lifted by the neuronal evolution. To see this, consider the early-time dynamics with uniform gains $p=\mathbf{1}/K$, for which
\[
\tau_x \dot x^i = -x^i + \frac{1}{N}\sum_{\mu=1}^{K}\xi_\mu^i\, m_\mu,
\qquad
m_\mu \equiv \sum_{j=1}^{N}\xi_\mu^j\,\phi(x^j).
\]
Differentiating $m_\mu$ gives $\dot m_\mu=\sum_i \xi_\mu^i\,\phi'(x^i)\dot x^i$, hence
\[
\tau_x \dot m_\mu
= -\sum_{i=1}^{N}\xi_\mu^i\,\phi'(x^i)x^i
+ \frac{1}{N}\sum_{\nu=1}^{K} m_\nu \sum_{i=1}^{N}\xi_\mu^i\,\phi'(x^i)\xi_\nu^i .
\]
Taking the difference between two overlaps yields
\begin{equation}
\tau_x\Delta \dot m
= -\sum_{i=1}^{N}\Delta \xi^i\,\phi'(x^i)x^i
+ \frac{1}{N}\sum_{\nu=1}^{K} m_\nu\sum_{i=1}^{N}\Delta \xi^i\,\phi'(x^i)\xi_\nu^i ,
\label{eq:breaking}
\end{equation}
where $\Delta \dot m = \dot m_\mu-\dot m_\rho$ and $\Delta\xi^i=\xi_\mu^i-\xi_\rho^i$. Except for nongeneric trajectories where the right-hand side vanishes identically, $\Delta \dot m\neq 0$ and the degeneracy is broken over time. Since $\dot f_\mu = (m_\mu/N)\dot m_\mu$, this induces a gain reweighting in the astrocytic dynamics. In practice, this mechanism implies that equal-fitness ties are generically transient and are resolved as $x(t)$ evolves away from the initialization.

\subsection{Fixed points of the system}
\label{app:fixed_points}

Given the system dynamics of Eq.\ref{eq:system_dynamics} fixed point conditions: 
\begin{align}
    & x^*=W(p^*) \phi(x^*)\\
    &p^*= \operatorname{softmax}\!\left(\frac{f(x^*)}{T} \right), 
\end{align}
where the second equation comes from Eq.\ref{eq:rep_fp}. Plugging the second equation into the first, it can be rewritten as a single equation on the $x$ domain. In this general regime, there are no analytic solutions, but we can see how the system behaves in different regimes: either hold the temperature fixed and take asymptotic limits of the time constants, or instead fix the time constants and examine the temperature limits.

\textbf{Case $\tau_p\to 0$:} In this case we instantly get $p^*_\mu = \text{softmax} \left(\frac{1}{2NT} \langle \xi_\mu, \phi(x) \rangle ^2 \right)$, so the evolution is written like: 
\[    \tau_x \frac{dx}{dt}  = - x + W(p) \,\phi(x)
\]
This corresponds to a biased Hopfield network in which every memory is weighted by $p^*_\mu$ like $W =\sum_\mu  p^*_\mu  \xi_\mu\xi_\mu^\intercal$. Since $p^*_\mu$ is highly heterogeneous when $|\langle \xi_\mu ,\phi\rangle |$ is high, we are biasing the Hopfield evolution matrix towards those patterns which are more likely to contain the pattern we are searching for. 

\textbf{Case $\tau_p\to \infty $:} If we take $\tau_p \to \infty$, the astrocytic modulation is frozen $\dot p=0$, that is, $p(t) = \bar p$ is constant. Since the astrocytic influence is initialized by the uniform vector $p = \mathbf{1}/K$, we recover the classical Hopfield dynamics 
\begin{equation}
\tau_x \dot x = - x + W(\bar p) \,\phi(x), 
\label{eq:Hopfield_regime}
\end{equation}
where with $\text{diag}(\bar p) =\text{diag}(1/K) $ and thus $\tau_x \dot x = -\alpha x + W \,\phi(x) $.

\textbf{Case $\tau_x\to 0$:} In this case, the neurons evolve instantaneously to their fixed point $x^*=W(p) \phi(x^*)$. This way, we have
\[    \tau_p \dot p =  (\text{diag} (p) - p p^\intercal) (f-T\log p),
\]
where $f_\mu=\frac{1}{2NT} \langle \xi_\mu, \phi(x^*) \rangle ^2$ is a transcendental function that depends also on $p$. Since the initial condition for the modulation equation is the uniform $p = \mathbf{1}/K$, the neurons initial configuration is already at a fixed point of the classical Hopfield dynamic $x^* = W(1/K) \phi(x^*)$. For a sufficiently large number of patterns, the initial fixed point will likely be a spurious minima. This means, in the very first instants of time, since $x(t_0)= x^*$, the fitness $f(x^*) $ of astrocytic modulation, will select those patterns that are the most similar to the $\phi(x^*)$ configuration, which can be arbitrarily far from $x_0$ and thus even becoming detrimental for retrieval. 

\textbf{Case $\tau_x \to \infty $:} Here we have $\dot x = 0$, meaning that $x(t)$ is constant it cannot evolve, so we stay forever in the initial condition even if the astrocyte dynamics converge to a fixed point in the $p$ domain.

\textbf{Case $T\to \infty$:} Keeping now the time scales constant, in this regime, the regularization effect on the s-replicator is so strong that the system stays in the $p = \mathbf{1}/K$ solution. Taking the astrocyte component, we can rearrange for the temperature to obtain
\[\frac{\tau_p }{T}\frac{d p_\mu}{d t} = \frac{1}{T}p_\mu (f_\mu-  \sum_{\nu}p_{\nu} f_\nu) -p_\mu (\log p_\mu-\sum_\nu p_\nu \log p_\nu).\]
Taking $T\to \infty$, we see that $\log p_\mu= \langle  \log p\rangle $ where we are forced into a maximum entropy distribution. This way we recover the classical Hopfield model.

\textbf{Case $T\to 0$:} Similarly, rearranging for temperature, if we take $T\to 0$, the dynamics become 
\[\tau_p \frac{d p_\mu}{d t} = p_\mu (f_\mu-  \sum_{\nu}p_{\nu} f_\nu), \]
in which case there is no regularization. The fixed points of the system are given by
\begin{align}
        & x^*=\frac{1}{\alpha }\Xi \text{diag}(p^*)\Xi^\intercal \phi(x^*)\\
        & p^* = \frac{\mathbf{1}_{ M(x^*)}}{|M(x^*)|}, 
\end{align}
where $M(x)$ is the set of $p_\mu$ elements that satisfy $M(x^*)=\arg\max_\mu f_\mu(x^*)$ and $\mathbf{1}_{ M(x^*)} \in \{0,1\}^K$. Note that $M(x^*)$ will generally have a single element unless in the contrived case discussed before in Eq.(\ref{eq:breaking}).

\subsection{Simulations details}
\label{app:simulation_details}

To integrate Eq.~(\ref{eq:dp}), we use an explicit Euler method. Trajectories are computed for $10/dt$ time steps, with $dt=0.001$, using smaller steps for small parameter value regimes (i.e. if $\alpha$ is the parameter then $dt = \alpha \cdot  0.05$ if $\alpha \leq  0.01$). For statistically significant results, we run $50$ simulations with different random pattern matrices $\Xi$ for every different parameter configuration. The observables are computed using the values of the last configuration values $x(t_f),p(t_f)$. To plot the percentile bands, we have smoothed out noise using a 1d Gaussian filter for better visualization. The code to reproduce the simulations is available at \cite{astrocytecode_github}. 

\bibliographystyle{apsrev4-2}
%\bibliography{bibliography}
%apsrev4-2.bst 2019-01-14 (MD) hand-edited version of apsrev4-1.bst
%Control: key (0)
%Control: author (72) initials jnrlst
%Control: editor formatted (1) identically to author
%Control: production of article title (-1) disabled
%Control: page (0) single
%Control: year (1) truncated
%Control: production of eprint (0) enabled
%

\end{document}